\documentclass[12pt]{article}

\usepackage[english]{babel}

\usepackage[a4paper]{geometry}

\usepackage{authblk}
\usepackage[colorlinks=true, allcolors=blue]{hyperref}

\usepackage{amsmath}
\usepackage{graphicx}

\usepackage{adjustbox}
\usepackage{graphicx}
\usepackage{tcolorbox}
\usepackage{graphicx}
\usepackage{empheq}

\usepackage{tikz}
\usetikzlibrary{math} 
\usepackage{import}
\usepackage{float}
\usepackage{xcolor}
\usepackage{amsmath}
\usepackage{multirow}
\usepackage{amsmath,amsfonts,amssymb,amsthm}
\usepackage{mathtools}
\usepackage[linesnumbered,ruled]{algorithm2e}
\usepackage{mathtools}
\usepackage[super]{nth}
\usepackage{caption}
\usepackage{subcaption}
\usepackage{graphicx}
\usepackage{wrapfig}
\usepackage{verbatim} 
\usepackage{longtable}
\usepackage{array}
\usepackage{supertabular}
\usepackage{longtable}
\usepackage{tabularx}
\usepackage{multirow}
\usepackage{colortbl}
\usepackage{fancyhdr}
\usepackage{xcolor}
\usepackage{pgf,tikz}
\usepackage{pgfplots, pgfplotstable}
\usetikzlibrary{fit,calc,positioning,decorations.pathreplacing,matrix}
\usetikzlibrary{arrows}
\usetikzlibrary{shapes}
\usetikzlibrary{chains}

\usepackage{multido}
\usepackage{calc}
\usepackage{fp}
\usepackage{etoolbox}
\DeclareMathAlphabet{\mathpzc}{OT1}{pzc}{m}{it}
\usepackage{breakcites} 
\usepackage[utf8]{inputenc}
\usepackage{helvet}
\usepackage{geometry}

\usepackage{amssymb}
\usepackage{enumitem}
\usepackage{tabularx}
\usepackage{xcolor}
\usepackage[absolute,overlay]{textpos}
\usepackage{graphicx}
\usepackage{lipsum}
\usepackage{caption}
\usepackage{multicol}
\usepackage{afterpage}
\usepackage{setspace}
\usepackage{pgffor}
\usepackage{parskip}

\usepackage{algpseudocode}

\algrenewcommand\algorithmicrequire{\textbf{Require:}}
\algrenewcommand\algorithmicensure{\textbf{Postcondition:}}
\algnewcommand\algto{\textbf{ to }}

\pgfplotsset{compat=1.18}

\newcommand{\SDvect}[1]{%
            \underline{\boldsymbol{\mathbf{%
                \mathit{#1}
            }}}
}

\newcommand{\SDtens}[1]{%
            \underline{\underline{\boldsymbol{\mathbf{%
                \mathit{#1}
            }}}}
}

\newcommand{\vect}[1]{\underline{{#1}}} 

\newcommand{\p}{\hspace{0.5mm}} 

\newcommand{\normi}[1]{\left\|{#1}\right\|}

\newcommand{\NS}{n_s} 

\title{A plausible Parametrization of Modal Basis for Dynamical Systems Analysis}

\author[1]{Sebastian Rodriguez}
\author[1]{Sergio Torregrosa}
\author[2]{Alicia Cordero}
\author[2]{Juan R. Torregrosa}
\author[1,3]{Mustapha Ziane}
\author[1,4]{Francisco Chinesta}

\affil[1]{PIMM Laboratory,
Arts et Métiers Institute of Technology, Paris, France.}

\affil[2]{Instituto de Matemática Multidisciplinar, Universitat Politècnica de València, Camino de Vera, s/n, 46022 Valencia, Spain.}

\affil[3]{ESI Group, 3bis, Rue Saarinen CEDEX, Rungis, 94528, France.}

\affil[4]{CNRS@CREATE LTD,
1 Create Way, \#08-01 CREATE Tower, Singapore.}

\begin{document}

\maketitle

\begin{abstract}

In the field of solid dynamics, knowing the corresponding modal basis of the system is capital, in order to improve design with respect to a desired dynamical behavior, such as avoiding natural frequencies at specific values or designing mechanical systems that can account for desired frequency spectrum. However, the determination of the modal basis involve the resolution of an eigenvalue problem, which can be expensive to perform for large systems, especially when dealing with a optimization of a parametric system design. In the present work, we propose to determine the parametrization of modal basis by considering an advanced Deep Learning technique based on the Rank Reduction AutoEncoder (RRAE). The RRAE is based on an autoencoder whose latent space is constrained through a truncated Singular Value Decomposition (SVD) approximation. This formulation enables the latent space to capture the dominant features of the data efficiently. As a result, the autoencoder is guided toward learning the underlying physical behavior represented across the dataset, mitigating overfitting and spurious predictions. The main idea consists of identifying a reduced parameter space using the RRAE for the first eigenvector, while the remaining modes are subsequently reconstructed through neural networks that take the same reduced parameter space as input, thereby coupling all modes in a nonlinear parametric framework. The proposed architecture is validated through the parametrization of the modal basis in 1D and 2D problems.

\end{abstract}

\section{Introduction}\label{sec:Introd}

The dynamical behavior of mechanical and structural systems is fundamentally governed by their modal properties. In particular, the modal basis, composed of the eigenvalues and eigenvectors of the governing equations, plays a central role in vibration analysis, structural optimization, and control-oriented design \cite{reddy1993introduction,allaire2018modal,li2020study}. Knowledge of the modal basis enables engineers to predict resonance phenomena, avoid critical natural frequencies, and tailor structures to achieve prescribed dynamic responses \cite{reddy2015introduction}. As a consequence, modal analysis is a cornerstone of solid dynamics and structural engineering.
For realistic engineering systems, modal properties are obtained by solving large-scale eigenvalue problems derived from discretized governing equations, typically using the Finite Element Method (FEM) \cite{reddy1993introduction}. While robust and accurate, these computations become increasingly expensive as system size grows or when repeated evaluations are required. This is particularly problematic in parametric design and optimization settings, where variations in geometry, material properties, or boundary conditions necessitate repeated eigenvalue solves. In such contexts, the computational cost of modal analysis can become a major bottleneck, motivating the development of efficient surrogate models capable of approximating modal quantities with reduced computational effort.

Model order reduction (MOR) techniques, such as Singular Value Decomposition (SVD) \cite{klema1980singular}, Certified Reduced Basis Method (CRBM) \cite{hesthaven2016certified}, Proper Orthogonal Decomposition (POD) \cite{liang2002proper} or Proper Generalized Decomposition \cite{chinesta2010recent,chinesta2011short,pruliere2010deterministic,ibanez2018multidimensional,sancarlos2021pgd}, have long been employed to alleviate this burden \cite{german2019reduced,sirkovic2016low}. These methods exploit the intrinsic low-dimensional structure of solution manifolds to approximate system behavior. Reduced-order techniques based on projection methods have been investigated for parameterized eigenvalue problems. In particular, Reduced Basis (RB) approaches construct low-dimensional approximation spaces from representative eigenvector snapshots in order to efficiently predict parameter-dependent eigensolutions. For example, \cite{fumagalli2016reduced} proposed a reduced basis framework equipped with a posteriori error estimators for parametrized elliptic eigenvalue problems, enabling accurate and computationally efficient approximation of eigenpairs over a parameter domain. Similarly, several POD- and RB-based approaches have been developed to approximate modal subspaces through offline-online decomposition strategies \cite{horger2017simultaneous,buchan2013pod}. However, these methods remain fundamentally based on linear reduced spaces, which may limit their ability to capture strongly nonlinear modal dependencies induced by complex parametric variations.

In recent years, Deep Learning techniques have emerged as powerful tools for data-driven surrogate modelling in computational mechanics. Neural networks have been successfully applied to approximate solutions of Partial Differential Equations (PDEs), reduced-order models, and parametric mappings between system inputs and outputs. Among these approaches, autoencoders have attracted significant attention due to their ability to learn low-dimensional latent representations of high-dimensional data \cite{goodfellow2016deep}. By encoding complex data structures into compact latent spaces, autoencoders provide a natural framework for nonlinear model reduction. Despite their potential, standard autoencoders may suffer from overfitting, lack of interpretability, and poor generalization when applied to physics-based problems. In particular, unconstrained latent spaces can lead to non-physical representations or hallucinated features that do not respect the underlying structure of the data.

In this work, we build upon these ideas and propose a novel Rank Reduction AutoEncoder (RRAE) \cite{mounayer2024rank} framework for the parametrization of modal basis in solid dynamics. The RRAE is based on an autoencoder whose latent space is constrained through a truncated Singular Value Decomposition (SVD) approximation. This formulation enables the latent space to capture the dominant features of the data efficiently. As a result, the autoencoder is guided toward learning the underlying physical behavior represented across the dataset, mitigating overfitting and spurious predictions. This technique has also been applied to a variety of problems, including its use as a Generative Design (GD) technique \cite{mounayer2025variational,tierz2025variational,idrissi2025generative,idrissi2025new}.

In this work, we extend the RRAE architecture to develop efficient nonlinear parametric surrogates. To this end, only the first eigenvector is learned directly by the RRAE, while its latent representation is simultaneously regressed using a neural network that takes the design parameters as input. Additionally, to couple all eigenvectors, $N$ supplementary neural networks are introduced, each receiving the latent space predicted by the parametric network and predicting one of the remaining $N$ eigenvectors. By training all networks jointly, the resulting framework yields a powerful parametric eigenvalue prediction model. This structured learning strategy exploits shared latent information across modes while preserving the flexibility required to capture mode-specific characteristics. The proposed methodology is demonstrated and validated on both a one-dimensional bar problem and a two-dimensional plate problem. The results highlight the potential of the RRAE framework as an efficient surrogate model for parametric modal analysis.

The paper is structured as follows. Section \ref{sec:RP} introduced the reference problem concerning dynamical systems and modal basis determination. Section \ref{sec:Param_RRAE} provides the details related to the RRAE and present the proposed learning strategy for modal basis parametrization. Section \ref{sec:numerical_example} illustrate the performance of the proposed architecture in a 1D bar and 2D plate numerical results. Finally, Section \ref{sec:conc_and_pers} provides Conclusions and Perspectives.

\section{Reference problem}\label{sec:RP}

In solid mechanics, the governing equations of motion are commonly obtained through finite element discretization, leading to the second-order semi-discretized system:
\begin{equation}
\SDtens{M} \p \ddot{\SDvect{u}}(t) + \SDtens{C} \p \dot{\SDvect{u}}(t) + \SDtens{K} \p \SDvect{u}(t) = \SDvect{f}(t) \ ,
\end{equation}
where $\SDtens{M}, \SDtens{C}, \SDtens{K} \in \mathbb{R}^{\NS \times \NS}$ denote the mass, damping, and stiffness matrices, respectively, $\NS$ is the number of spatial degrees of freedom (DoFs) of the system, and $\SDvect{u}(t) \in \mathbb{R}^{\NS}$ represents the vector of nodal displacements at time $t$.

In the absence of external forces and damping, free vibration analysis yields the generalized eigenvalue problem:
\begin{equation}\label{eq:Eigenvalue_prob}
\SDtens{K} \p \SDvect{\phi}_i = \omega_i^2 \SDtens{M} \p \SDvect{\phi}_i, \qquad \forall i = 1,\ldots,\NS
\end{equation}
where $\omega_i$ is the natural frequency of eigenmode $i$ and $\SDvect{\phi}_i \in \mathbb{R}^{\NS}$ the corresponding mode shape.

In this context, eigenvalues $\omega_i^2$ represent the intrinsic stiffness of the structure, while eigenvectors $\SDvect{\phi}_i$ describe the spatial distribution of displacement associated with each vibration mode. These quantities provide a direct physical interpretation of how mass and stiffness interact to produce dynamic response.


\subsection{Parametric Modeling and the need for Surrogate Modeling}

From a design perspective, eigenvalues and eigenvectors are fundamental performance indicators. Natural frequencies determine susceptibility to resonance and dynamic instability, while mode shapes identify regions of high displacement, strain, and stress.

Design modifications, such as changes in material properties, geometry, or boundary conditions, alter the system matrices $\SDtens{M}$ and $\SDtens{K}$, thereby modifying the spectral properties of the structure. As a result, eigenvalues and eigenvectors become implicit functions of a set of design parameters $\vect{p}$:
\begin{equation}
\omega_i = \omega_i(\vect{p}), \quad \SDvect{\phi}_i = \SDvect{\phi}_i(\vect{p}), \qquad \forall i = 1,\ldots,\NS
\end{equation}

This parametric dependence is central to vibration-based design objectives, such as frequency maximization, mode separation, or dynamic compliance minimization.

The repeated solution of large-scale eigenvalue problems in design optimization can be computationally prohibitive. To address this challenge, parametric surrogate models are often constructed to approximate the dependence of eigenvalues and eigenvectors on design parameters. This will be the main subject of the following section.

\section{Parametric Surrogate model construction}\label{sec:Param_RRAE}

In this section, we address the construction of the parametric surrogate model. However, before that, one needs to address a very important step, which consists on data cleaning and preparation. This data preparation is of key importance when approximating eigenvectors, this preparation consists on \textit{mode matching} and \textit{alignment}.

The eigenvectors of a mechanical system are determined up to a scaling factor, meaning each mode shape has two solutions of opposite signs (positive or negative). This can be easily check by introducing $\SDvect{\phi}_i$ or $-\SDvect{\phi}_i$ in equation \eqref{eq:Eigenvalue_prob}. Also, when computing the eigenvectors for different system parameters $\vect{p}$, the ordering of the modes can be changed. The previous mentioned characteristic of eigenvectors can completely perturb the correct learning process of Machine Learning algorithm if not addressed correctly.

In this sense, in the following section, one address these problems as a necessary pre-processing step before feed it up to the parametric RRAE to produce the surrogate model.

\subsection{Pre-processing data: Re-ordering and alignment of eigenvectors}


The first goal is to identify pairs of modes of the same system, disturbed and nominal ones. A system can be considered disturbed if, for instance, a perturbation of the system or a modification of its parameters $\vect{p}$ have been introduced. Moreover, the disturbed modes can appear in a different order and flipped with respect to the nominal ones when numerically computed for this new parameter set. Thereafter, the nominal modes matrix is noted $\SDtens{\Phi}(\vect{p})$ and the disturbed one $\SDtens{\hat{\Phi}}(\vect{p})$. Once the pairs are identified, we can re-order and align the disturbed eigenvectors with respect to the nominal ones. 

Here, many techniques exists to achieve this objective, however, as a matter of simplicity, we propose the strategy presented in Section \ref{sec:L1}.

\subsubsection{L1 matching technique}\label{sec:L1}

This problem is first addressed by a L1-regularized minimization approach. The L2 norm between the nominal mode and the corresponding disturbed one would be minimum. Since we seek for a bijective matching between disturbed and nominal modes, a L1 penalization is added to enhance sparsity. Hence, for each nominal mode $j$ we obtain a sparse vector of weights $\SDvect{\beta}^{j} \in \mathbb{R}^{\NS}$. The closest value to $1$ (or $-1$) in $\SDvect{\beta}^{j}$ indicates the corresponding disturbed mode $k$ for the nominal mode $j$:
\begin{equation}\label{eq:eig_order_problem}
	\forall j = 1,\ldots,\NS, ~~ \SDvect{\beta}^{j} = \underset{  \SDvect{\beta}^{j}  
 }{\text{arg min}} \ \| \SDvect{\phi}_{j}(\vect{p}) - \SDtens{\hat{\Phi}}(\vect{p})\SDvect{\beta}^{j} \|^{2}_{2} + \lambda \sum_{k} \vert\beta_{k}^{j}\vert \ ,
\end{equation}
where $\lambda$ corresponds to a regularization factor, that drives the degree of sparsity impose in the resolution. This problem is solved using the Lasso solver \cite{xu2008robust}.

\subsubsection{Numerical example of pre-processing data}

\begin{figure}[H]
	\centering
	\begin{subfigure}{0.45\textwidth}
		\includegraphics[width=\textwidth]{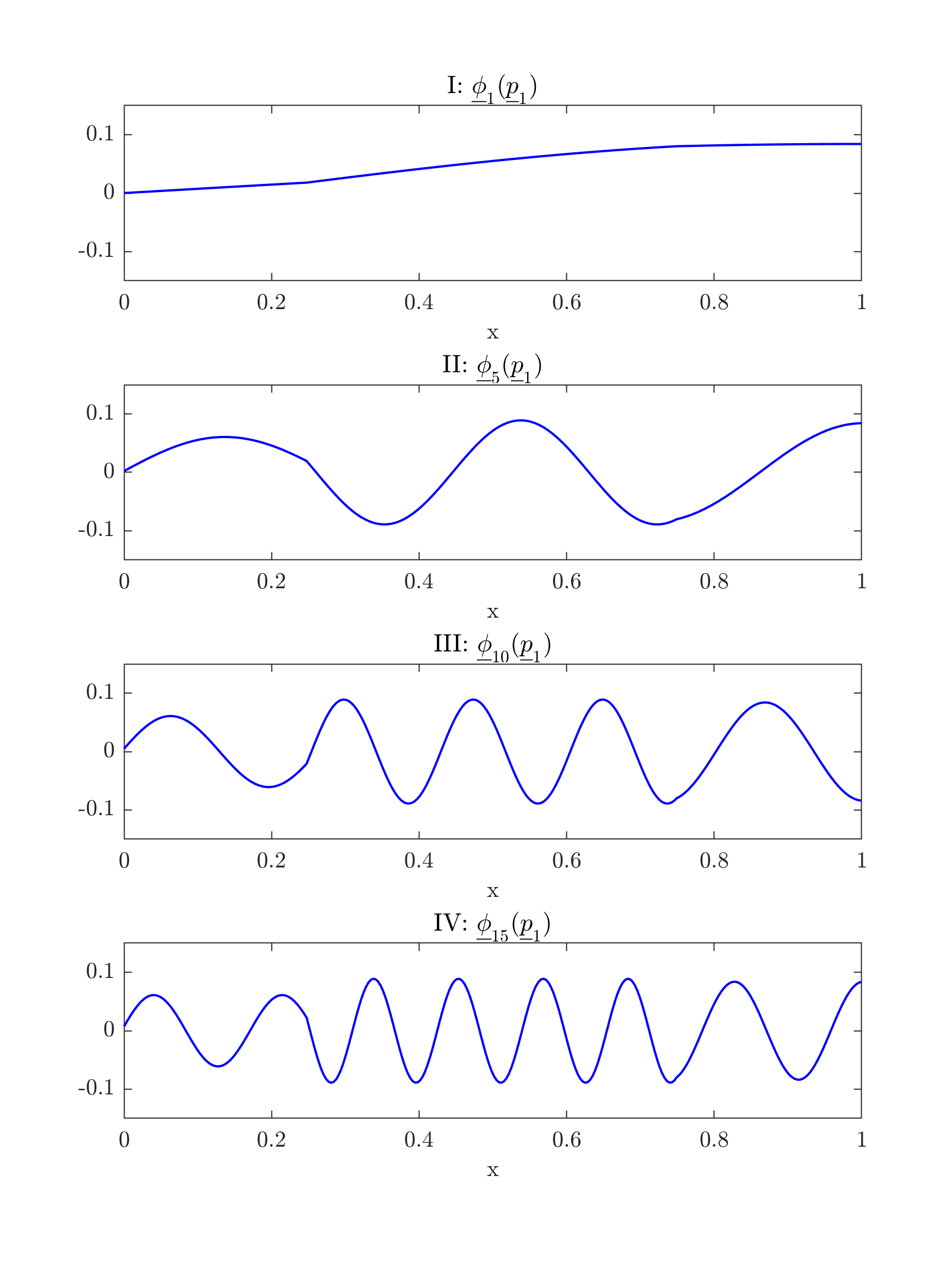}
		\caption{Nominal modes.}
		\label{OriginalModes}
	\end{subfigure}
	\hfill
	\begin{subfigure}{0.45\textwidth}
		\includegraphics[width=\textwidth]{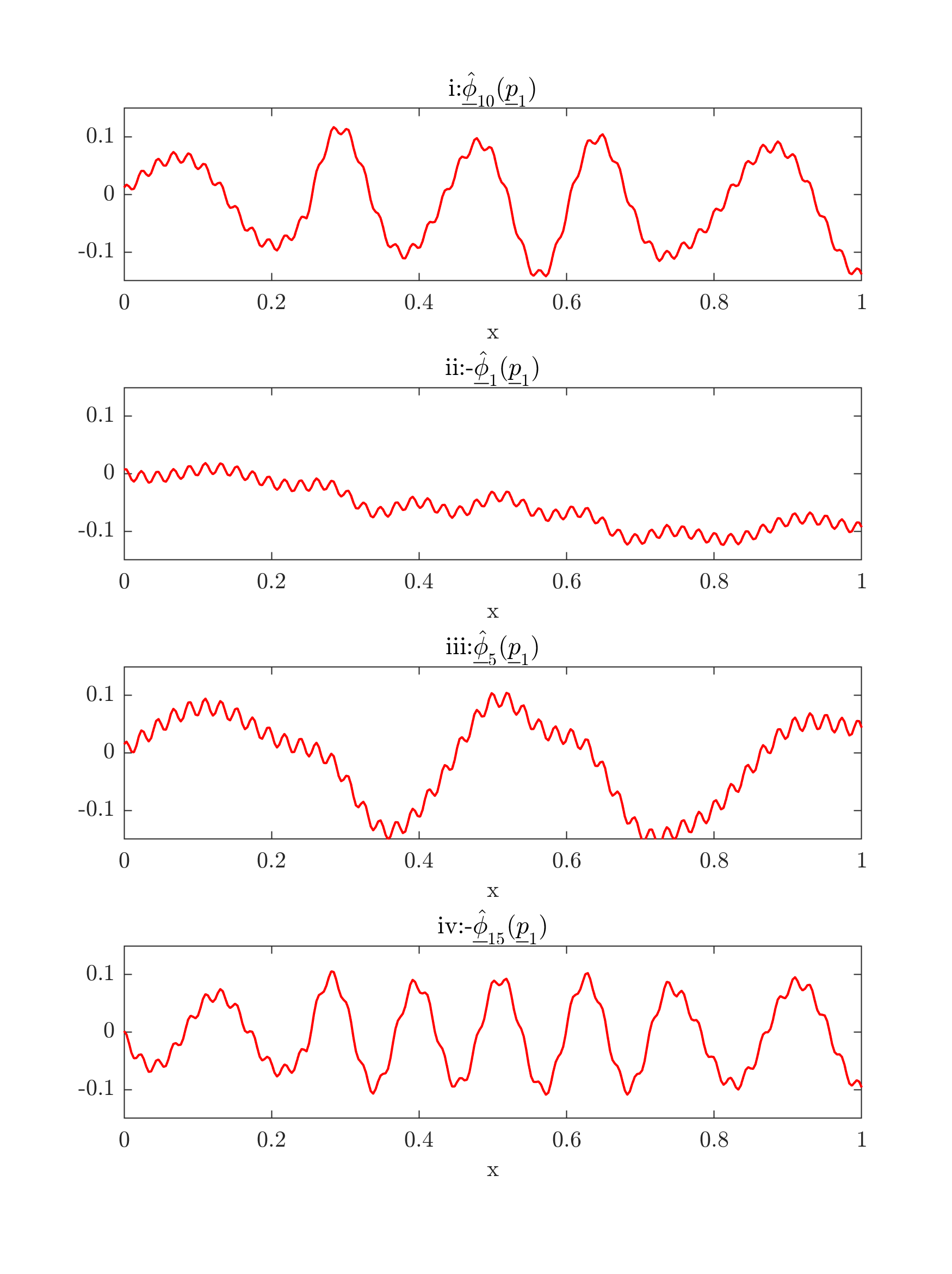}
		\caption{Noisy and disordered modes.}
		\label{DisturbedModes}
	\end{subfigure}	\caption{Nominal and disturbed modes.}
	
		\label{OriginalDisturbedModes}
\end{figure}

To illustrate the aforementioned method, here we artificially introduced a disturbance in the system. In order to obtain a disturbed mode, we first take the corresponding nominal mode and randomly multiply it by $1$ or $-1$. Then, local noise is added by a sinusoidal function of frequency $f_0$. In this context, we have:
\begin{equation}
	\SDvect{\hat{\phi}}_{i}(\vect{p}) = \pm \SDvect{\phi}_{i}(\vect{p}) + A\sin(f_0 \times \SDvect{x}) \ ,
\end{equation}
where $A \in \mathbb{R}$ and $\SDvect{x} \in \mathbb{R}^{\NS}$ corresponds to the amplitude of imposed noise and the spatial coordinate in $1$D respectively. The nominal and disturbed modes are presented in Figure \ref{OriginalDisturbedModes}.

\begin{figure}[H]
	\centering
	\begin{subfigure}{0.45\textwidth}
		\includegraphics[width=\textwidth]{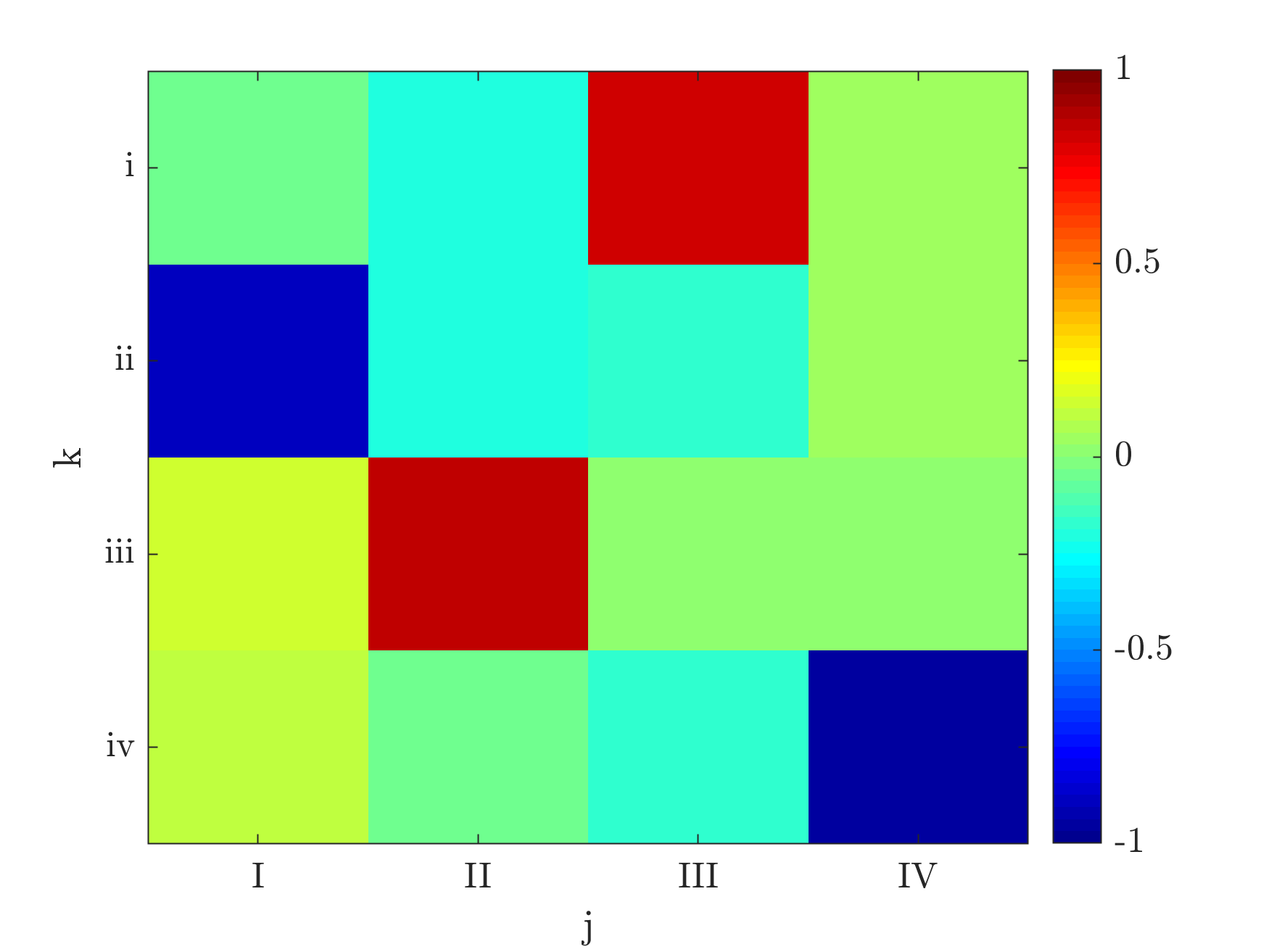}
		\caption{$\beta_{k}^{j}$ values obtained.}
		\label{L1Matrix}
	\end{subfigure}
	\hfill
	\begin{subfigure}{0.45\textwidth}
		\includegraphics[width=\textwidth]{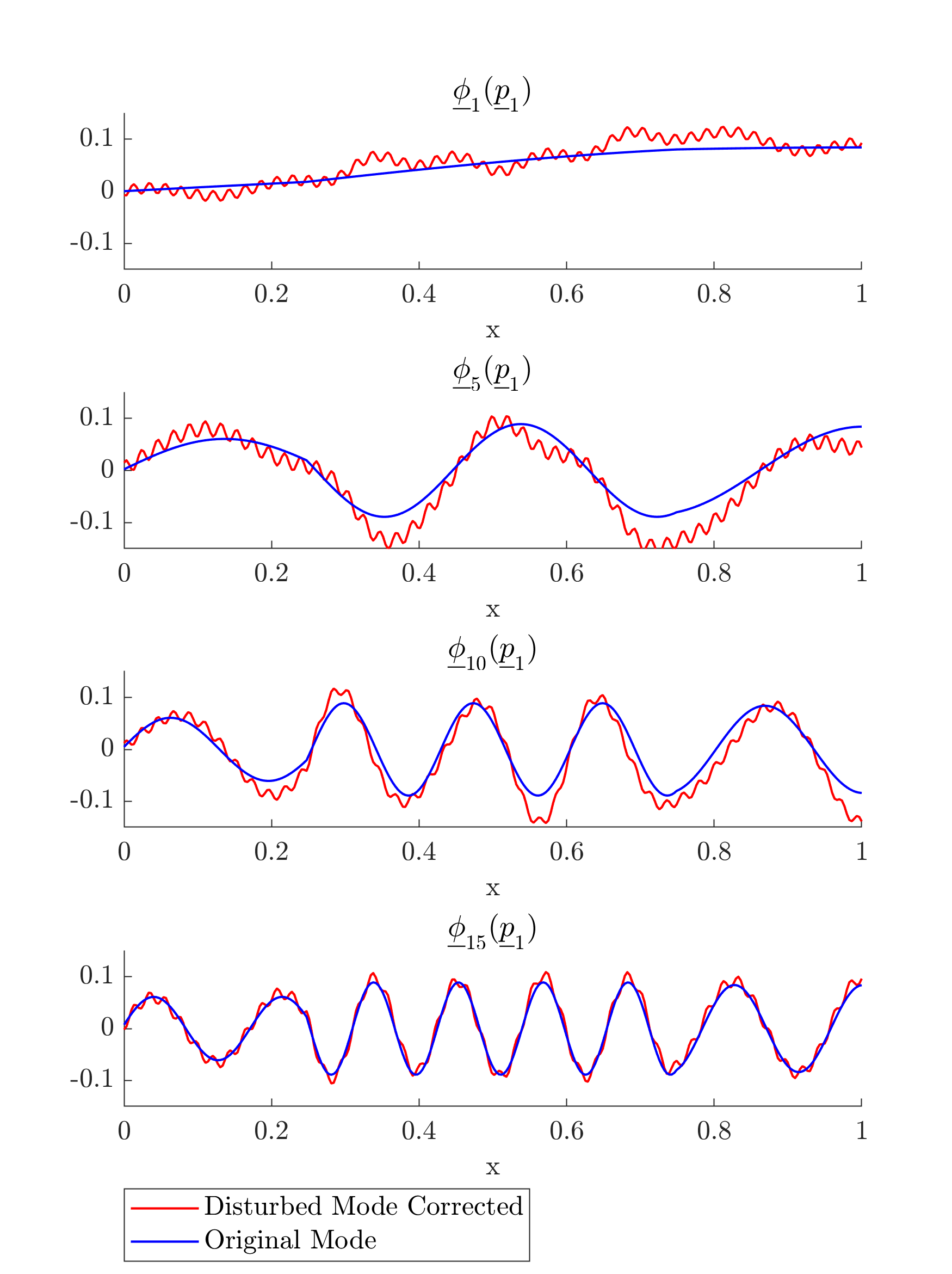}
		\caption{Aligned and matched modes.}
		\label{L1CorrectedModes}
	\end{subfigure}
	\caption{ Figure \ref{L1Matrix} illustrate $\beta_{k}^{j}$ factor when solving problem \eqref{eq:eig_order_problem} and Figure \ref{L1CorrectedModes} shows ordered and flipped modes.}
	\label{L1Results}
\end{figure}

The computed weights $\SDvect{\beta}^{j}$ for each nominal mode $j$ are presented in Figure \ref{L1Matrix}. The sparsity of the obtained values can easily be observed leading to a straightforward matching between nominal and disturbed modes. Moreover, the sign of this value indicates if the disturbed mode is flipped or not. Then, the disturbed modes can be re-ordered and flipped if needed, as illustrated in Figure \ref{L1CorrectedModes}.

\subsection{The Rank Reduction AutoEncoder for the parametrization of Eigenvectors}\label{sec:RRAE_eigv}


The proposed parametric surrogate model is built upon the Rank Reduction AutoEncoder (RRAE) \cite{mounayer2024rank}, a deep learning architecture specifically designed to extract compact and informative representations from the treated data. The objective is to obtain a latent representation that captures the dominant features of the data. To achieve this, the method integrates a low-rank latent space constraint.

The RRAE extends the traditional autoencoder structure by imposing a low-rank constraint on the latent space. Given an input signal $\SDvect{X} \in \mathbb{R}^T$, the encoder network $\mathcal{E}(\cdot)$ maps it into a latent vector $\SDvect{Y} \in \mathbb{R}^L$. Instead of allowing this latent vector to take an arbitrary form, the RRAE restructures it into a matrix whose Singular Value Decomposition (SVD) is truncated to a predefined rank $k_{\text{max}} \ll L$. In this regard, when doing training, and considering a batch strategy for this, one has:
\begin{equation}
\SDvect{Y} \approx \SDtens{V} \ \SDvect{\alpha} \ ,
\end{equation}
in where $\SDtens{V} \in \mathbb{R}^{L \times k_{\text{max}}}$ denotes the SVD basis, and $\SDvect{\alpha} \in \mathbb{R}^{k_{\text{max}}}$ the reduced variables.

This constraint ensures that the latent space captures only the most significant components of the encoded signals. Finally, the decoder $\mathcal{D}(\cdot)$ reconstructs the original signal from this low-rank latent representation. During training, the autoencoder optimizes the reconstruction error, pushing the encoder to learn latent variables that efficiently represent the data.

\begin{figure}[h!]
\centering
\includegraphics[width=1\textwidth]{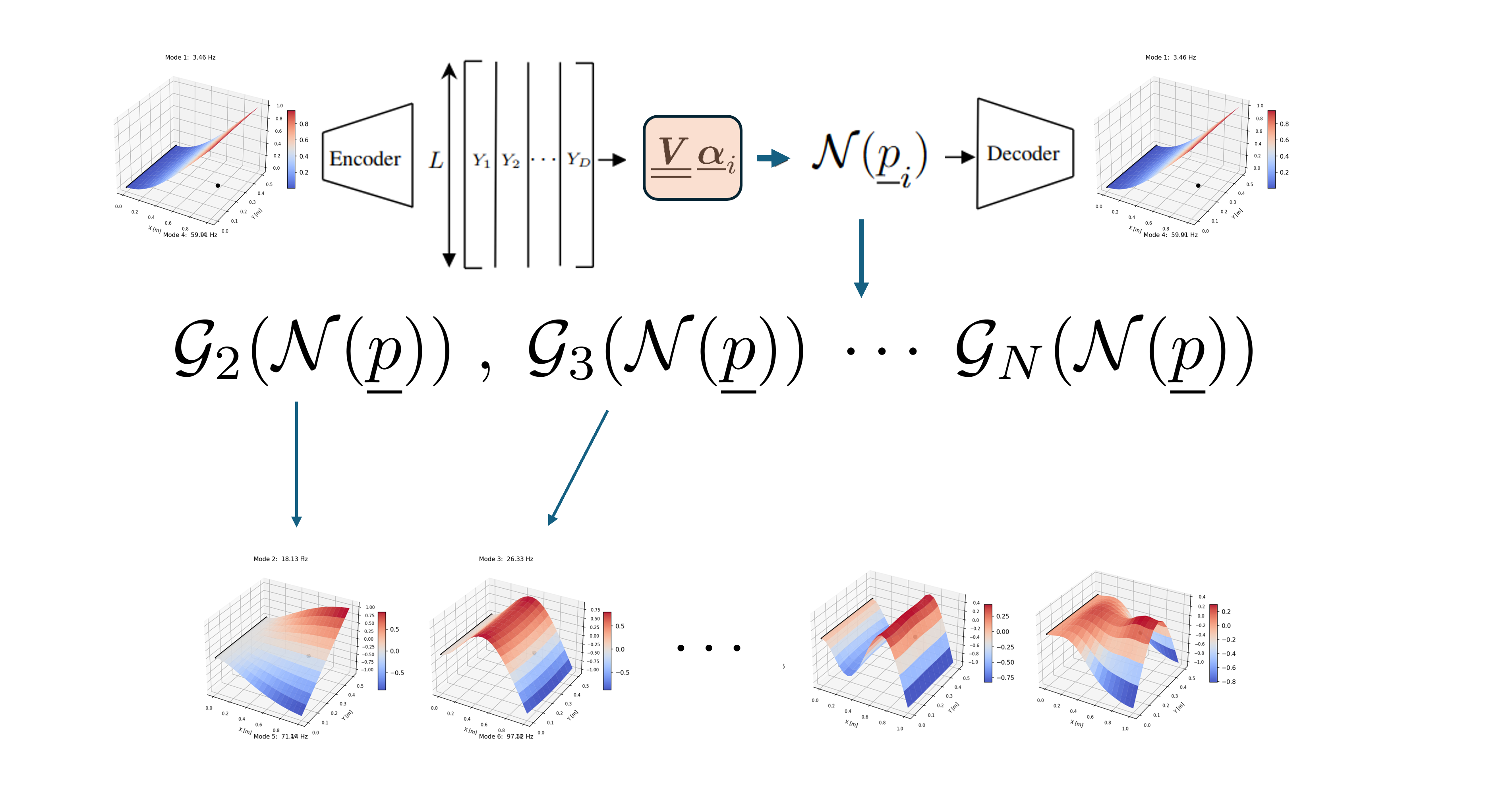}
\caption{Parametric surrogate model architecture.}
\label{fig:RRAE_param_eigv}
\end{figure}

The low-rank constraint of the RRAE helps to isolate the dominant features of the treated data. However, this by itself does not allow the parametrization of the eigenvectors. To achieve this, we introduce an additional neural network that takes the design parameters $\vect{p}$ as input and predicts the low-rank latent representation of the RRAE associated with the first eigenvector. Furthermore, to parameterize the remaining eigenvectors, we employ $N-1$ supplementary neural networks. Each network $i$ receives as input the prediction of the parametric network---namely, the reduced latent representation corresponding to the first mode (i.e $\SDvect{X}(\vect{p}) = \SDvect{\phi}_1(\vect{p})$)---and predicts the $i^{\text{th}}$ eigenvector of the system, with $i>1$. In this way, the complete set of eigenvectors, or any selected subset of $N-1$ additional modes, can be reconstructed. The overall architecture is illustrated in Figure \ref{fig:RRAE_param_eigv}.

%
%
By jointly training the RRAE along with the parametric layer as well as the Networks related to reproduce each of the eigenvectors, the latent space is shaped not only to minimize reconstruction error but also to maximize its relevance to perform eigenvectors predictions. 


From the previous explanation, the total loss function combines three main components:

\begin{itemize}
    \item \textbf{Reconstruction loss} $\mathcal{L}_\text{rec}$, encouraging accurate recovery of the original signals:
%
%
%
%
\begin{equation}
\mathcal{L}_\text{rec} = 100 \times \frac{ \normi{\mathcal{D}(\mathcal{N}(\vect{p})) - \SDvect{\phi}_1(\vect{p}) }_2 }{ \normi{\SDvect{\phi}_1(\vect{p}) }_2 } \ .
\end{equation}

    \item \textbf{Parametric Network loss} $\mathcal{L}_\text{param}$, link parameters design $\vect{p}$ to the regularized low-rank latent space of the RRAE.
\begin{equation}
\mathcal{L}_\text{param} = 100 \times \frac{ \normi{\mathcal{N}(\vect{p}) - \SDtens{V} \p \SDvect{\alpha}(\vect{p}) }_2 }{ \normi{\SDtens{V} \p \SDvect{\alpha}(\vect{p})}_2 } \ .
\end{equation}

    \item \textbf{Eigenvectors prediction loss} $\mathcal{L}_\text{eigv}$, guiding the latent representation in order to be able to predict all the other eigenvectors:
\begin{equation}
\mathcal{L}^{[N]}_\text{eigv} = \sum_{i=2}^{N} 100 \times \frac{ \normi{ \mathcal{G}_{i}(\mathcal{N}(\vect{p})) - \SDvect{\phi}_i(\vect{p}) }_2 }{ \normi{\SDvect{\phi}_i(\vect{p}) }_2 } \ ,
\end{equation}

\end{itemize}
with $\normi{\bullet}_2$ the Euclidean norm.

So, the total loss is given as:
\begin{equation}\label{eq:total_loss}
\mathcal{L} = \mathcal{L}_\text{rec} + \mathcal{L}_\text{param} + \mathcal{L}^{[N]}_\text{eigv} \ .
\end{equation}

\textbf{Remark}: Here, one could impose an additional penalization term imposing the orthogonality with respect to the Mass matrix \cite{reddy1993introduction}, as eigenmodes should verify. However, this imposition in practice did not improve the prediction of the parametric surrogate, so here it is not considered.

\section{Numerical examples}\label{sec:numerical_example}

Here, we demonstrate the performance of the proposed parametric surrogate model on both one-dimensional and two-dimensional solid mechanics problems. This is considered for simplification purposes, however, eigenvectors from 3D solid mechanics problems follows the same methodology and therefore are omitted in this work.

\subsection{1D bar}\label{sec:1D_test_case}

Here, the one-dimensional bar is assumed to have a length of $L = 1 [\text{m}]$ and to be isotropic, with Young’s modulus $E = 200 \  [\text{GPa}]$ and density $\rho = 1600 \ [\text{Kg}/\text{m}^3]$.

\begin{figure}[H]
\centering
\includegraphics[width=0.5\textwidth]{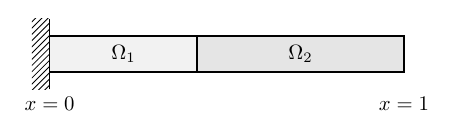}
\caption{Reference 1D bar considered.}
\label{fig:Reference_bar}
\end{figure}

\begin{figure}[H]
\centering
\includegraphics[width=0.5\textwidth]{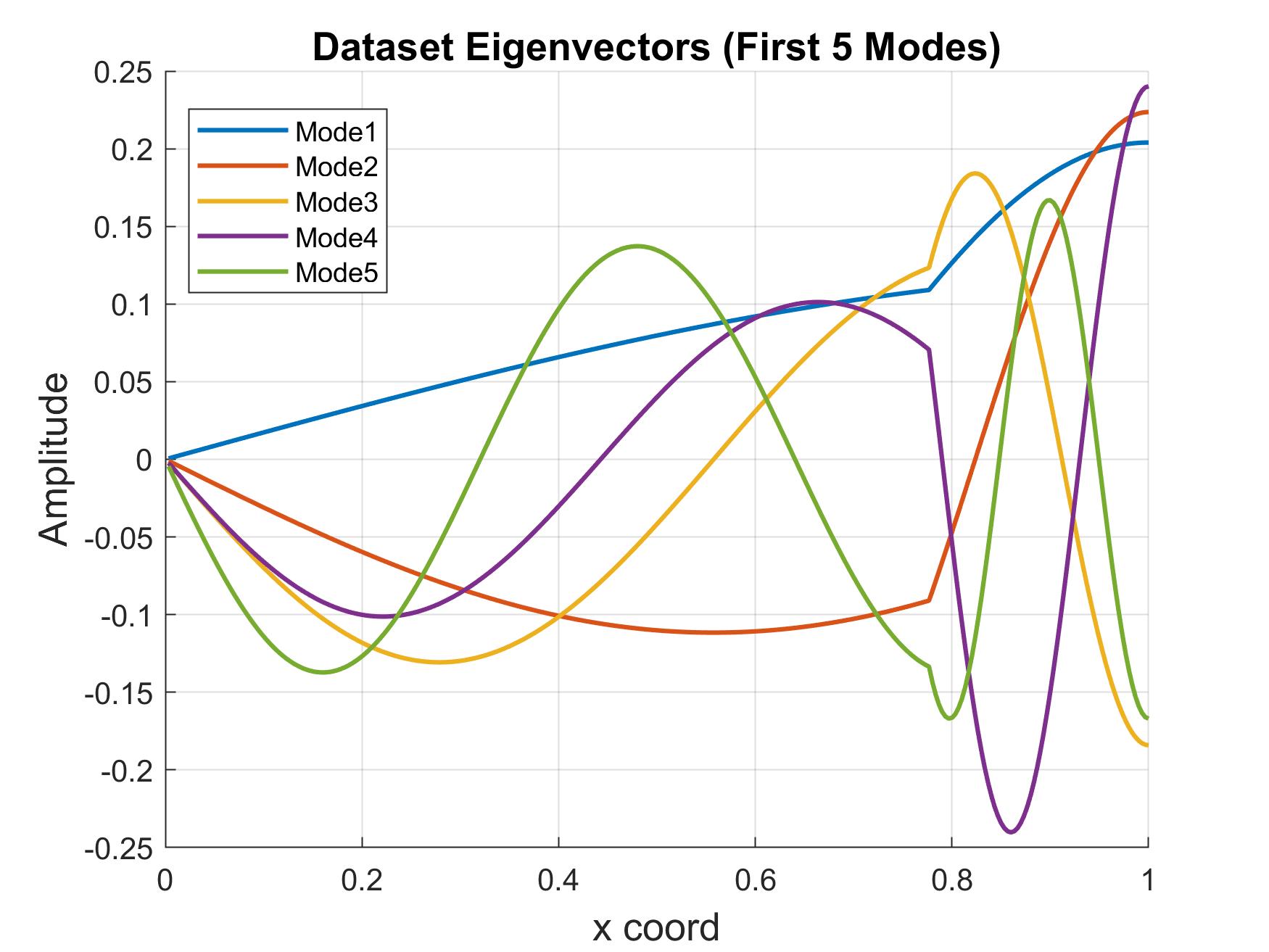}
\caption{Group of 5 Eigenvectors for a given fraction $g$.}
\label{fig:Dataset}
\end{figure}

We consider the bar composed of two different materials defined over two distinct spatial domains, whose relative length is characterized by the parameter $g$, such that:
\begin{equation}\label{eq:fraction_1D}
g = \frac{\Omega_1}{\Omega_2} \in [0,1] \ .
\end{equation}
The considered bar is illustrated in Figure \ref{fig:Reference_bar}.
Figure \ref{fig:Dataset} illustrate the first $5$ eigenvectors for this configuration when considering a given fraction of materials $g$.

Therefore, we consider as parameter of the problem the fraction $g$ of equation \eqref{eq:fraction_1D}. To construct a parametric surrogate model for the eigenvectors, a Design of Experiments (DoE) is employed, in where $2000$ combinations of the fraction $g$ were considered using Latin Hyper Cube \cite{loh1996latin} sampling. Then, we separate this dataset on train and test, where $80 [\%]$ correspond to train and $20 [\%]$ to test.

\begin{figure}[h!]
	\centering
	\begin{subfigure}{0.45\textwidth}
\includegraphics[width=\textwidth]{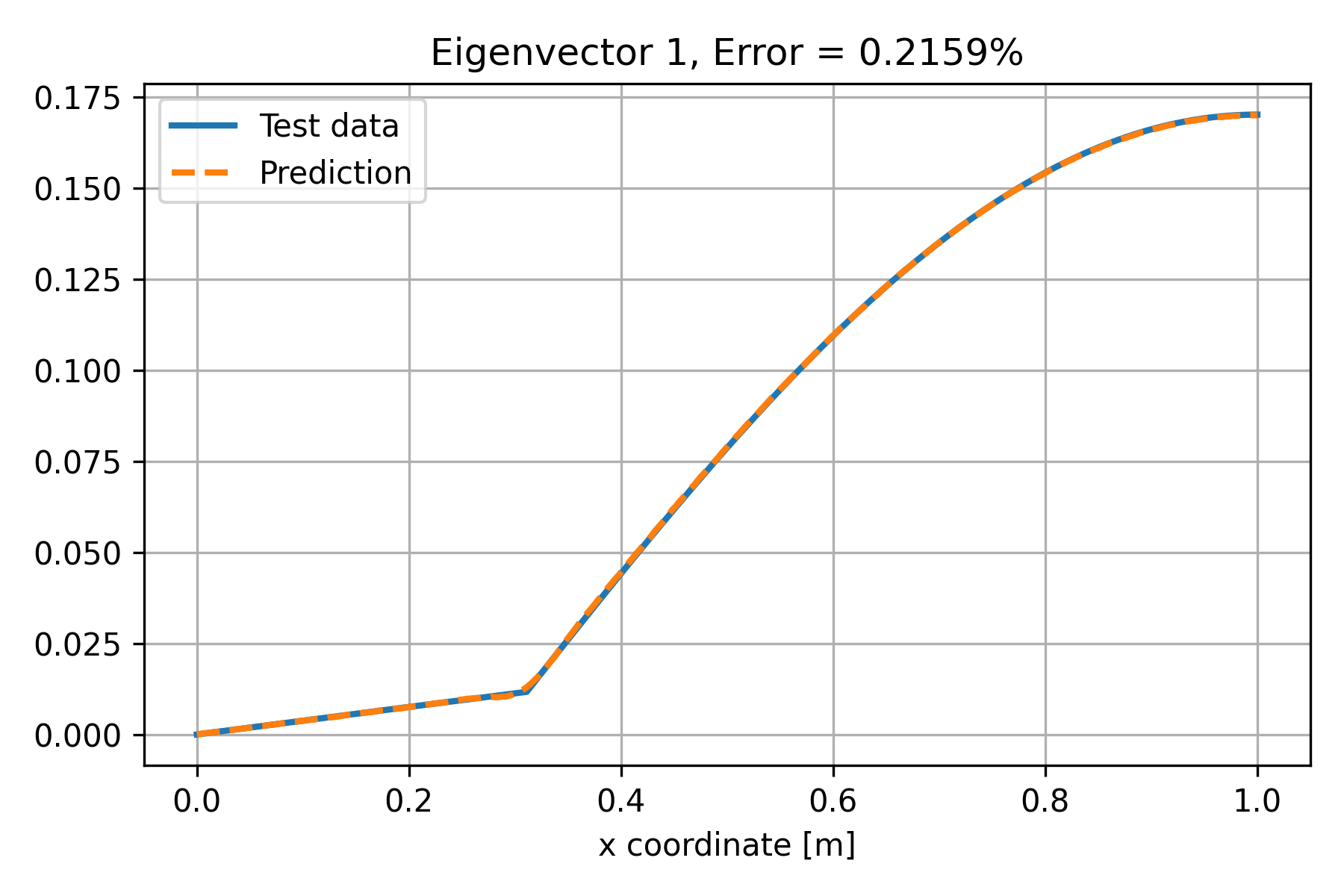}
		\caption{First eigenvector.}
		\label{fig:RRAE_19_mode_1}
	\end{subfigure}
	\begin{subfigure}{0.45\textwidth}	\includegraphics[width=\textwidth]{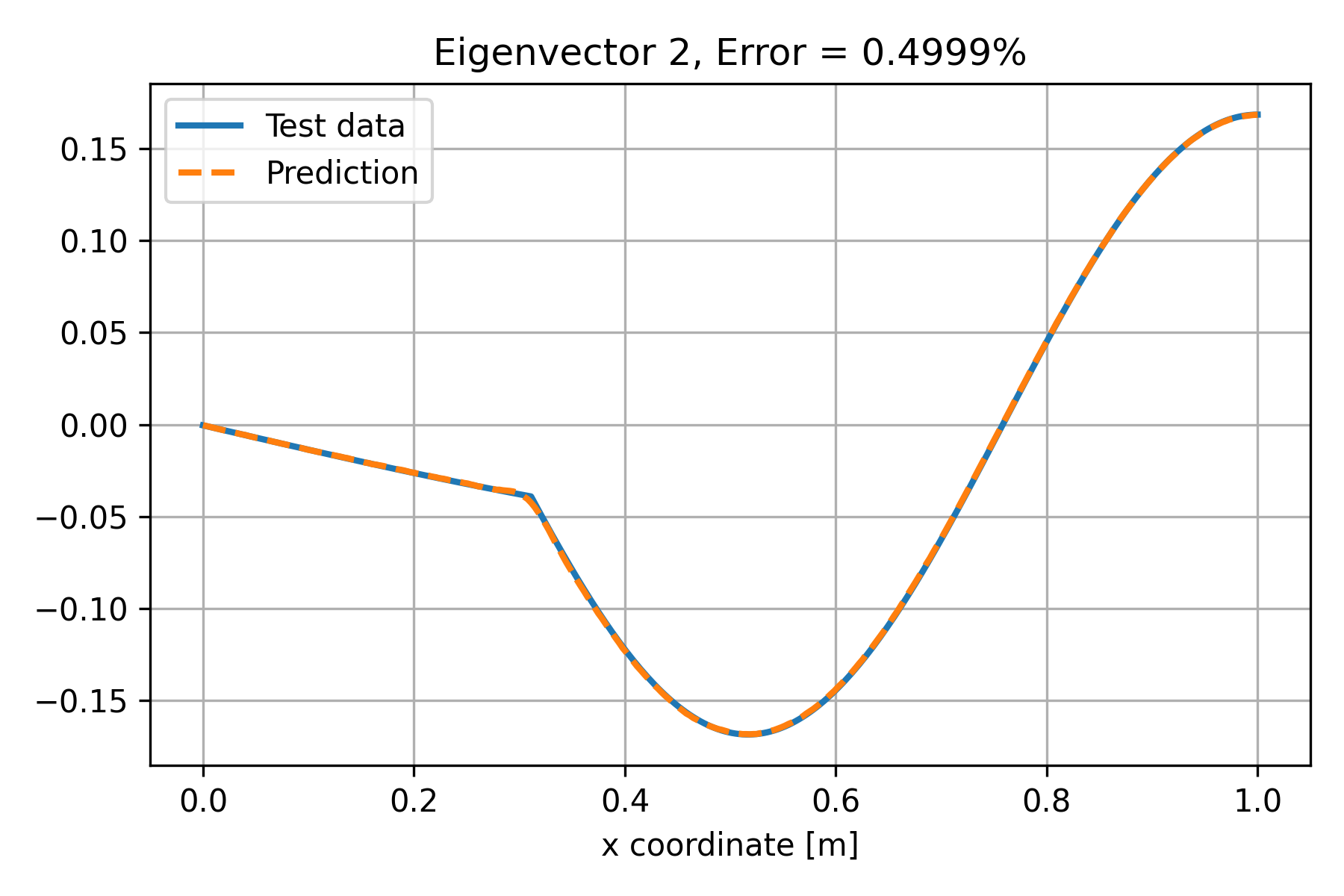}
		\caption{Second eigenvector.}
		\label{fig:RRAE_19_mode_2}
	\end{subfigure}
\begin{subfigure}{0.45\textwidth}	\includegraphics[width=\textwidth]{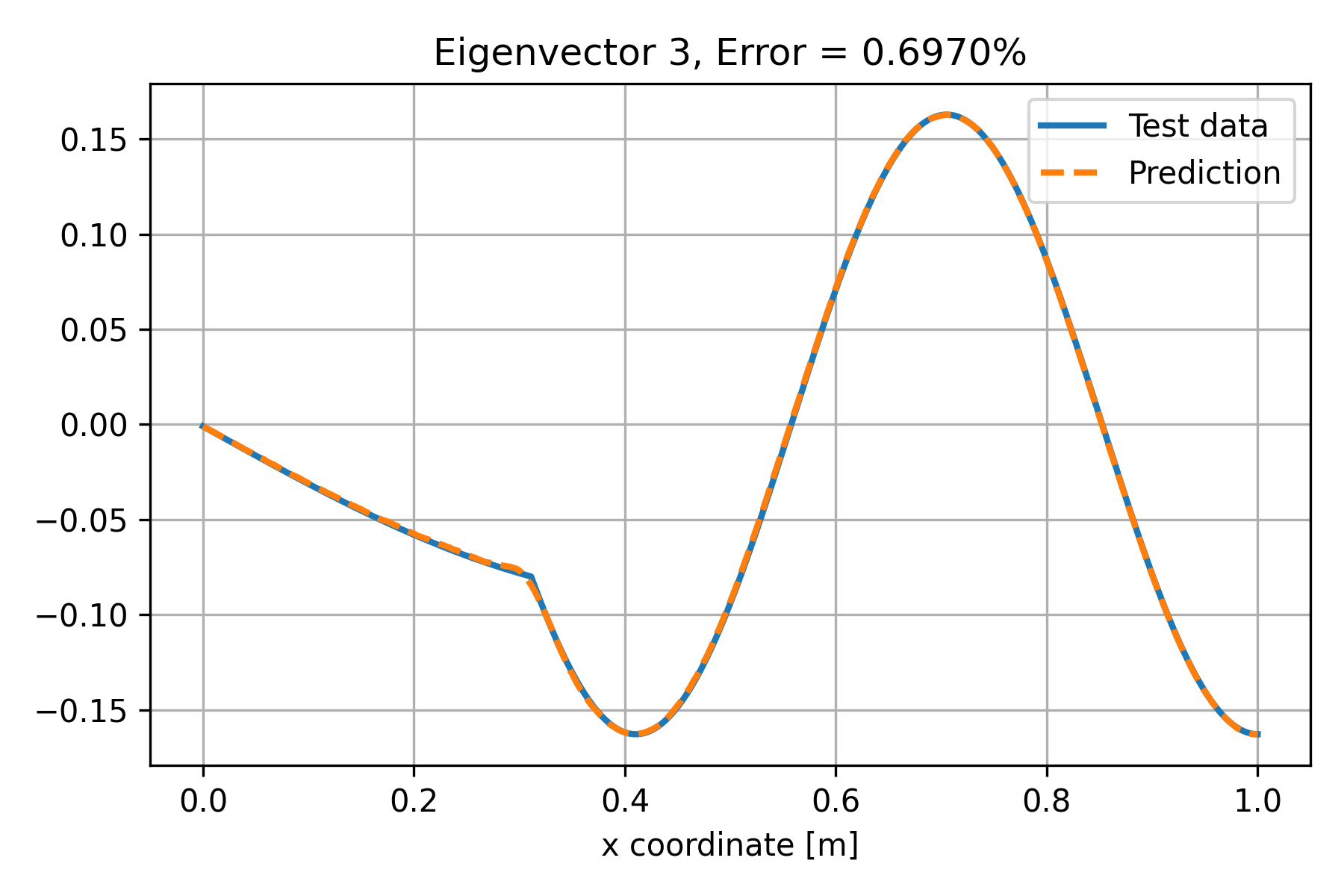}
		\caption{Third eigenvector.}
		\label{fig:RRAE_19_mode_3}
	\end{subfigure}
\begin{subfigure}{0.45\textwidth}	\includegraphics[width=\textwidth]{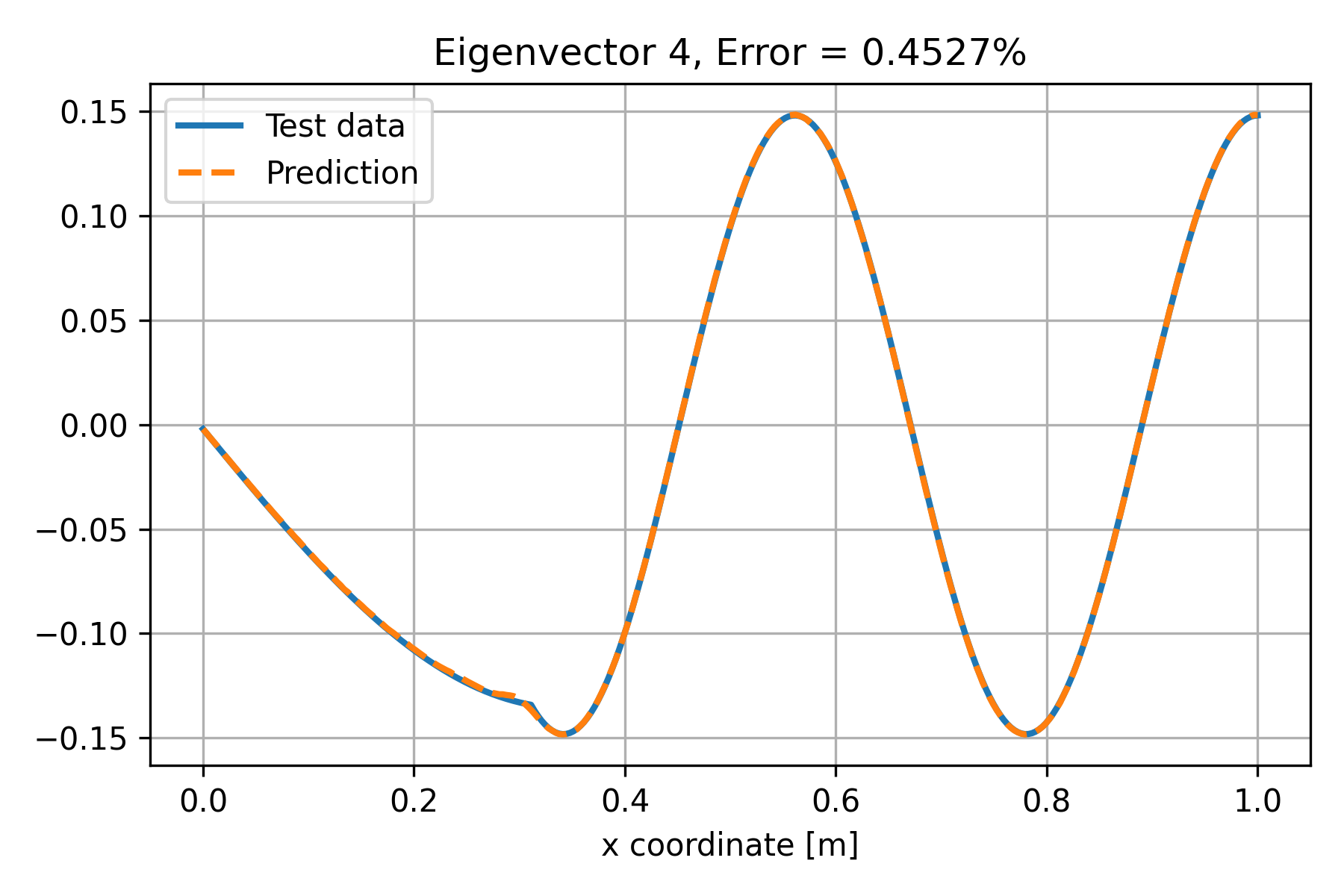}
		\caption{Fourth eigenvector.}
		\label{fig:RRAE_19_mode_4}
	\end{subfigure}	
\begin{subfigure}{0.45\textwidth}	\includegraphics[width=\textwidth]{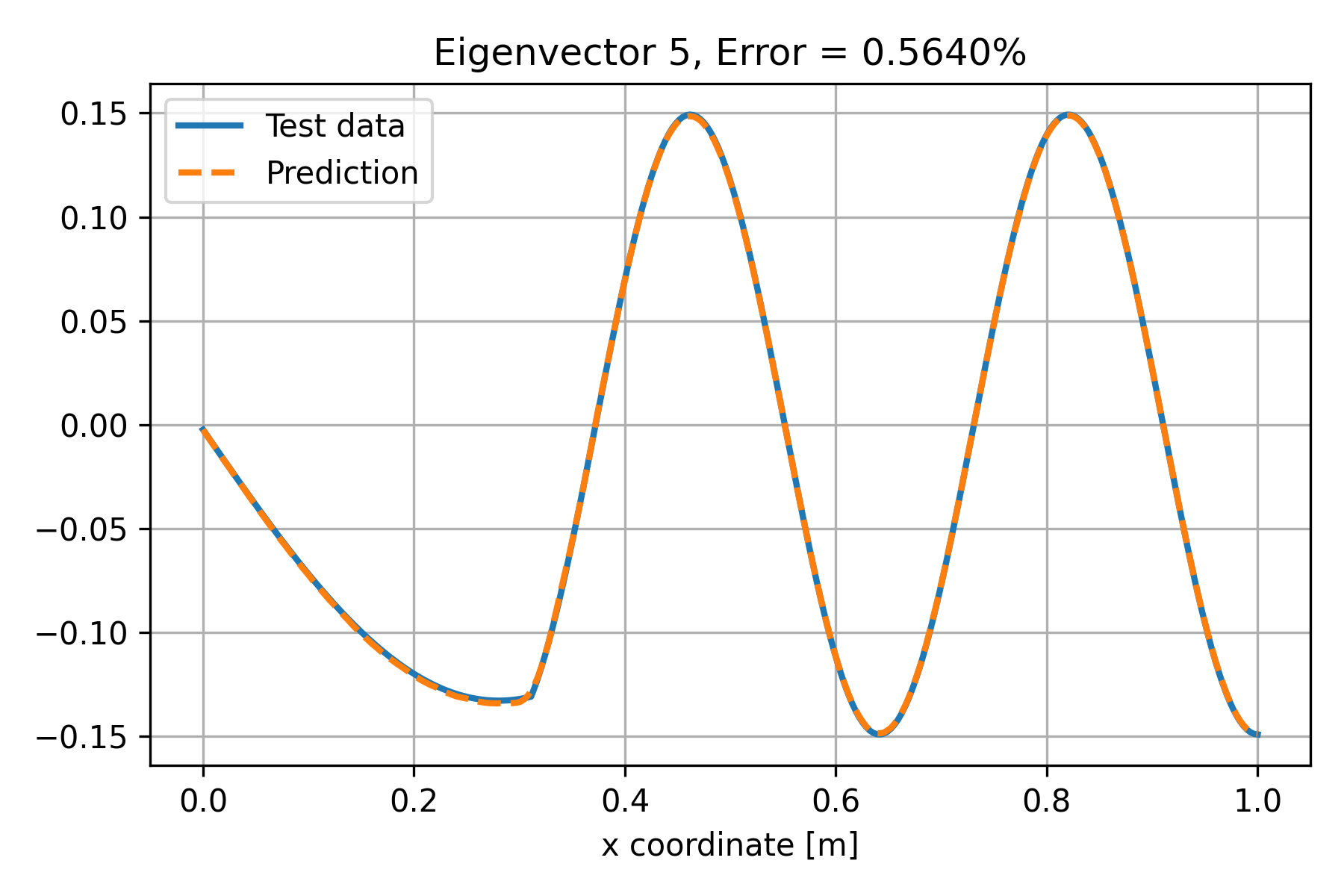}
		\caption{Fifth eigenvector.}
		\label{fig:RRAE_19_mode_5}
	\end{subfigure}		
		\caption{Parametric RRAE predictions of eigenvectors for the 1D bar at a given test data.}	\label{fig:1D_bar_RRAE_pred}
\end{figure}

To train the parametric surrogate model one considers $k_{\text{max}} = 1$ since only one physical parameter is considered (the fraction of domain $g$). Figure \ref{fig:1D_bar_RRAE_pred} illustrate the prediction of the parametric surrogate for a test dataset for the group of $5$ eigenmodes considered.

From previous results, one can clearly see that the group of eigenvectors selected to be approximated is correctly predicted by the parametric surrogate, certifying the quality of the surrogate.

\begin{figure}[H]
\centering
\includegraphics[width=0.8\textwidth]{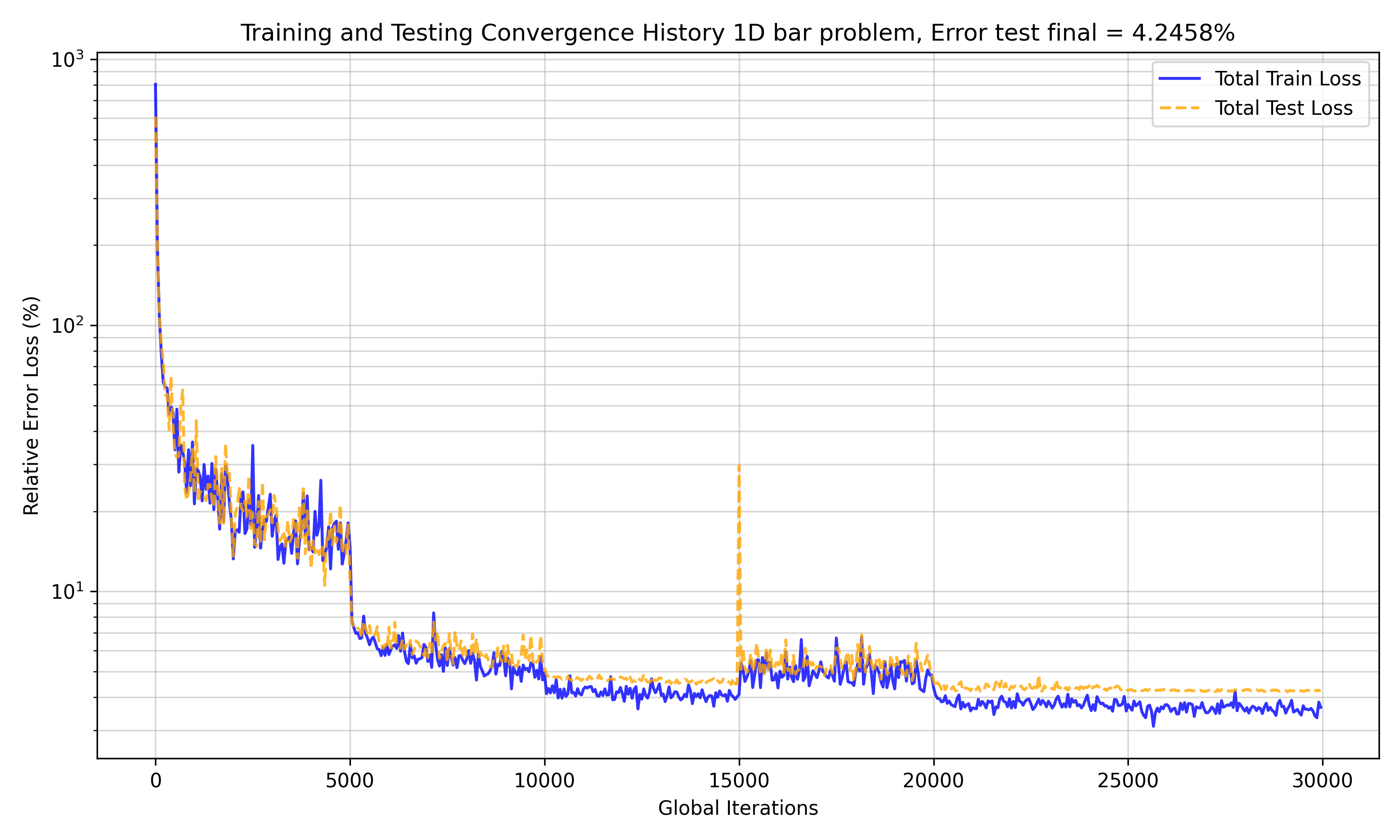
}
\caption{Train and Test loss in function of training iterations.}
\label{fig:Error_loss_bar}
\end{figure}

The train and test loss computed from equation \eqref{eq:total_loss} is illustrated in Figure \ref{fig:Error_loss_bar}.

\subsection{2D bending plate}\label{sec:2D_test_case}

Here, we consider the structural dynamics of a thin, flat, rectangular cantilever plate governed by the Kirchhoff-Love plate theory. The plate occupies a continuous spatial domain $\Omega = [0, L] \times [0, W] \subset \mathbb{R}^2$, where $L = 1 [\text{m}]$ denotes the length along the $x$-axis and $W =  0.5 [\text{m}]$ denotes the width along the $y$-axis. The plate features a uniform thickness $h = 5 [\text{mm}]$, mass density $\rho = 7800 \ [\text{Kg}/\text{m}^3]$ and Poisson's ratio $\nu = 0.3$. The plate is rigidly clamped along the edge $x = 0$ and remains entirely free along the other three boundaries ($x = L$, $y = 0$, and $y = W$). 

\begin{figure}[h!]
\centering
\includegraphics[width=0.8\textwidth]{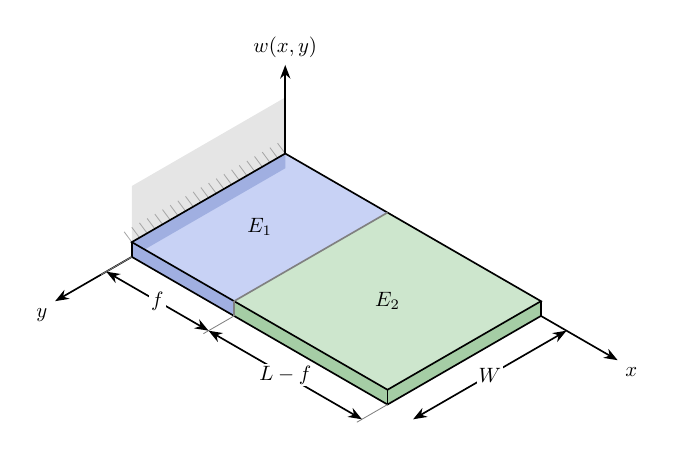}
\caption{Reference 2D plate considered.}
\label{fig:2D_plate}
\end{figure}

To evaluate the sensitivity of the system’s eigenspace, we consider the plate as divided into two domains, with a fraction denoted by f, where each domain is assigned a different Young’s modulus, $E_1$ and $E_2$ respectively, as illustrated in Figure \ref{fig:2D_plate}.
\begin{equation}
    \vect{p} = \begin{bmatrix} g & E_1 & E_2 \end{bmatrix}^T \in \mathcal{P} \ ,
\end{equation}
with \(g \in [0,1]\), \(E_1 \in [0.7,\,2.1]\,\text{GPa}\), and \(E_2 \in [70,\,210]\,\text{GPa}\).

For this numerical example, we considers the first $4$ eigenmodes and $500$ data points, where $80 [\%]$ correspond to train and $20 [\%]$ to test. 

To train the parametric surrogate model we considered $k_{\text{max}} = 3$ since the eigenvectors depends only on $3$ physical parameters. Also, for the encoder and decoder, as well as the additional neural networks that predicts the rest of the modes here we considered 2D Convolutional Neural Networks.

\begin{figure}[H]
\centering
\includegraphics[width=0.7\textwidth]{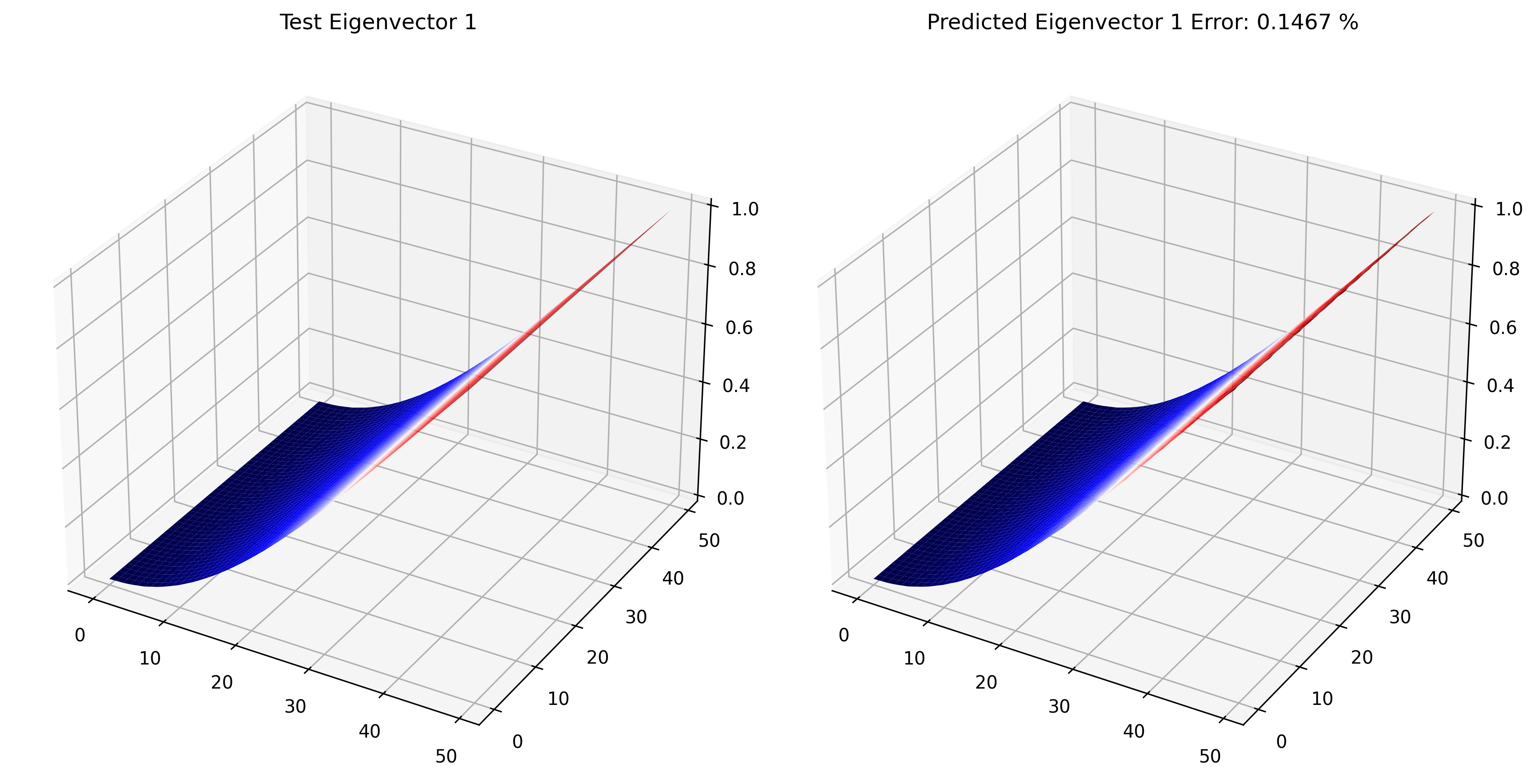}
\caption{Reference and prediction of first eigenvector.}
\label{fig:Plate_mode_1}
\end{figure}

\begin{figure}[H]
\centering
\includegraphics[width=0.7\textwidth]{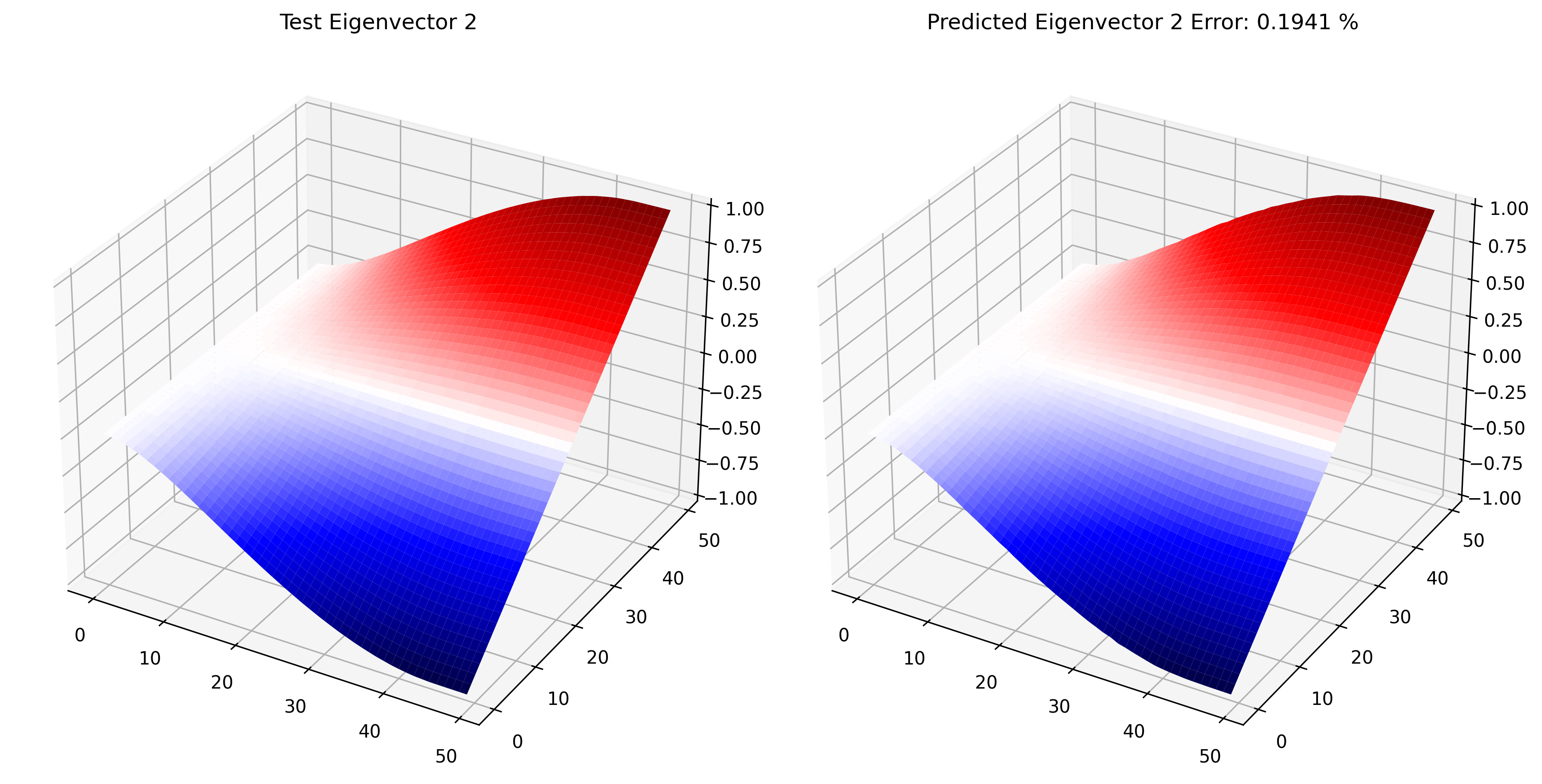}
\caption{Reference and prediction of second eigenvector.}
\label{fig:Plate_mode_2}
\end{figure}

\begin{figure}[H]
\centering
\includegraphics[width=0.7\textwidth]{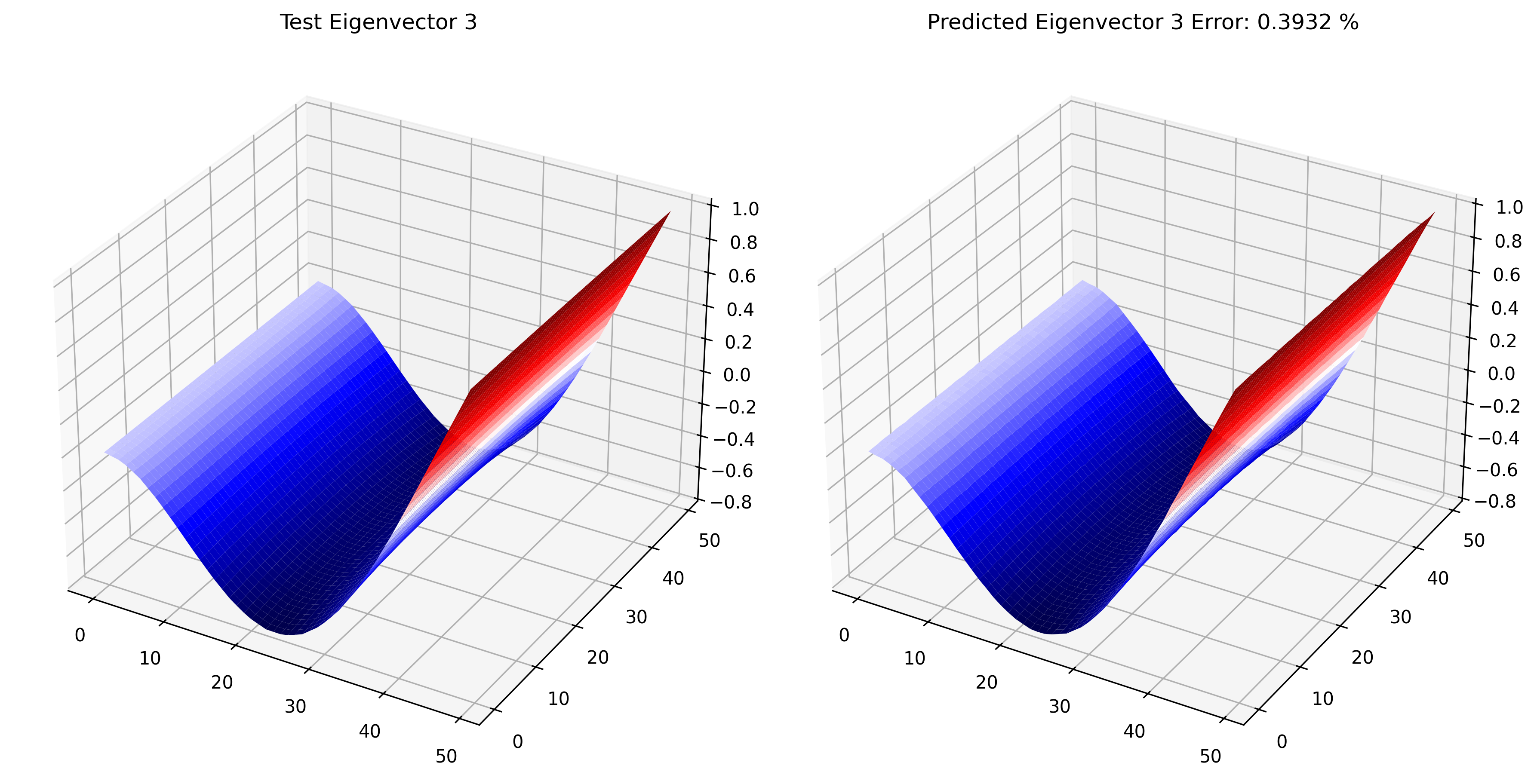}
\caption{Reference and prediction of third eigenvector.}
\label{fig:Plate_mode_3}
\end{figure}

Figures \ref{fig:Plate_mode_1}, \ref{fig:Plate_mode_2} and \ref{fig:Plate_mode_3} illustrate the reference and prediction of the parametric surrogate for a test dataset for eigenmodes $1$, $2$ and $3$.

\begin{figure}[H]
\centering
\includegraphics[width=0.7\textwidth]{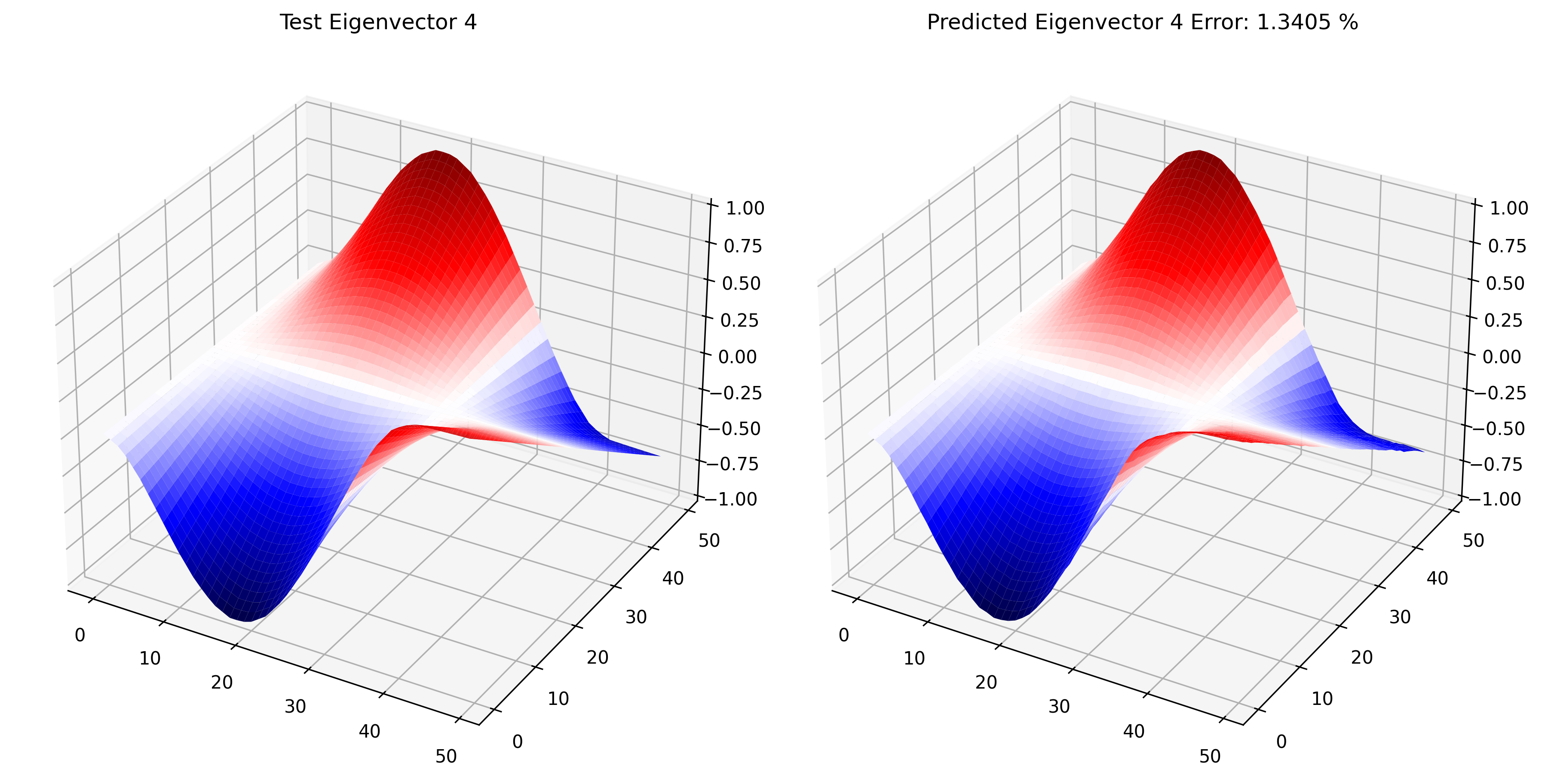}
\caption{Reference (left) and prediction (right) of the $4$th eigenvector for the dataset $20$.}
\label{fig:Plate_mode_4_1}
\end{figure}

\begin{figure}[H]
\centering
\includegraphics[width=0.7\textwidth]{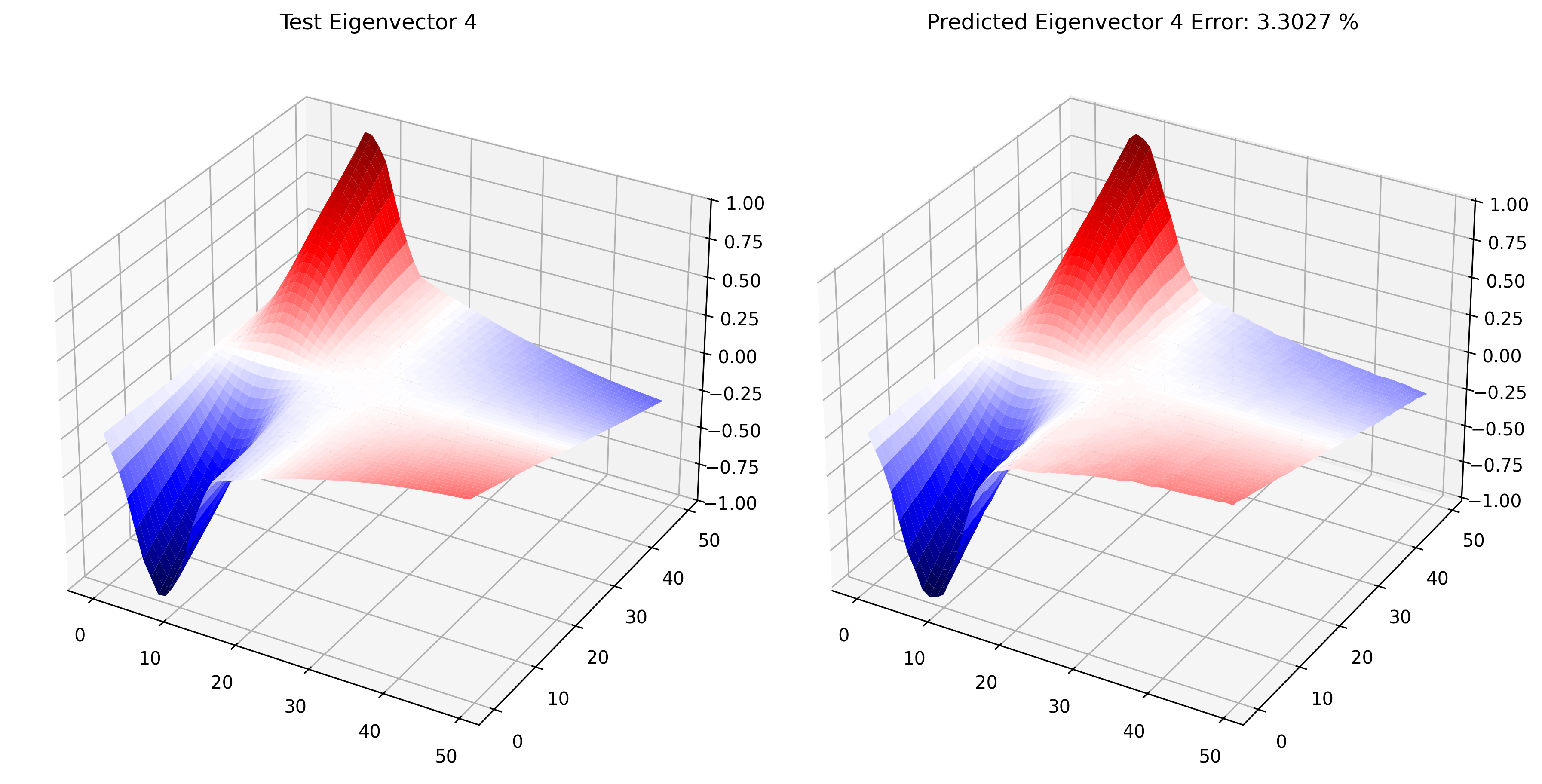}
\caption{Reference (left) and prediction (right) of the $4$th eigenvector for the dataset $40$.}
\label{fig:Plate_mode_4_2}
\end{figure}

The $4$th mode is more affected by the different stiffness defined on the two domains. Figure \ref{fig:Plate_mode_4_1} and \ref{fig:Plate_mode_4_2} shows the reference and prediction of the fourth mode for two test cases. 

\begin{figure}[H]
\centering
\includegraphics[width=0.8\textwidth]{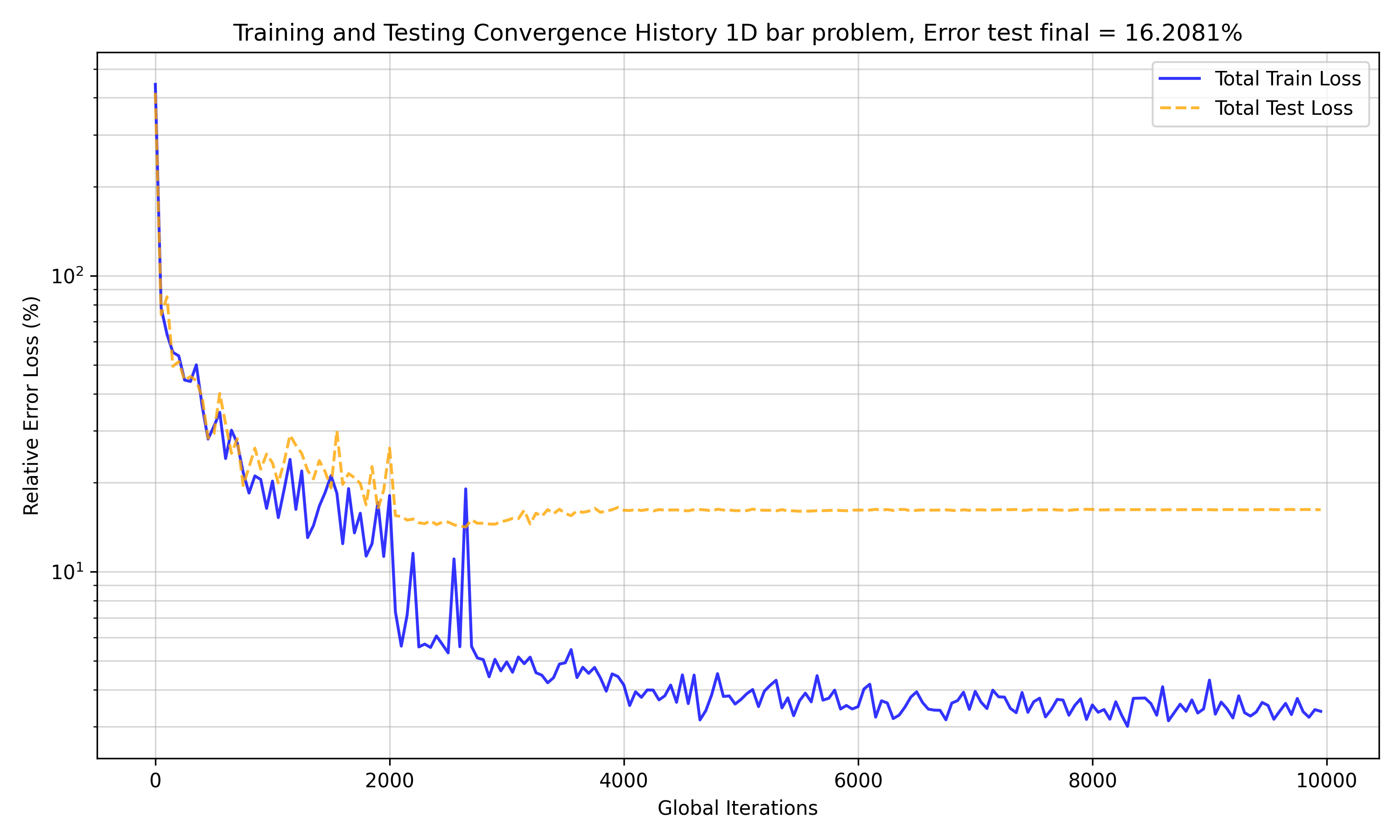
}
\caption{Train and Test loss in function of training iterations.}
\label{fig:Error_loss_plate}
\end{figure}

Finally, the train and test loss computed using \eqref{eq:total_loss} is illustrated in Figure \ref{fig:Error_loss_plate}.
It should be noted that the error definition, which coincides with the loss function used in the RRAE architecture, is obtained by summing the errors associated with all eigenvectors. Consequently, a test error of $16$ [\%], although seemingly high, does not represent the error of each mode individually. In fact, the error associated with each mode on the test dataset remains below $4$ [\%].

\section{Conclusions and perspectives}\label{sec:conc_and_pers}

This work proposed a data-driven algorithm based on the nonlinear model-order reduction technique Rank Reduction AutoEncoder (RRAE) in order to address the parametrization of eigenvectors, a topic of relevance for solid dynamics, especially for system optimization design.

The RRAE consists of an autoencoder in where its latent space is constraint to be represented as a low-rank SVD approximation. This ingredient, allows to capture the most fundamental features on the data in order to reproduce them. Here, we went even further, by applying the additional constraint of being able to reproduce correctly the left group of eigenvectors of the eigenmodes for a given parameters set $\vect{p}$.

The parametrization is achieved using a neural network that takes the design parameters as input and predicts the low-rank latent space of the first eigenvector. The additional eigenvectors are then predicted through a set of $N$ distinct neural networks, each approximating a single eigenvector while sharing the same low-rank latent representation predicted by the parametric regressor. By training all the Networks simultaneously as well as the RRAE, at convergence these ingredients brings into place a robust and powerful architecture able to accurately predict eigenvectors over the parametric space.

As perspective we considers the treatment of 3D solid mechanics problems, nevertheless the proposed methodology will remains the same.

\bibliographystyle{unsrt}

\bibliography{sample}

\end{document}